\documentclass[journal]{IEEEtran}
\usepackage{cite}
\usepackage{graphicx,amssymb,lineno}
\usepackage{amsmath,amsfonts,amssymb,bm}
\usepackage{mathrsfs}

\usepackage{algorithm}
\usepackage{algorithmic}
\usepackage[usenames]{color}
\usepackage{float}
\usepackage{mathtools}
\usepackage{booktabs}
\usepackage{bm}

\usepackage{graphicx,graphics,color,epsfig,subfigure,graphpap,rotate}
\usepackage{times, verbatim, subfigure, epsfig, graphicx, latexsym, amsmath}
\usepackage{url}
\usepackage{subfigure}
\usepackage{CJK}
\usepackage{makecell}
\begin{document}

\title{Joint User Association and Transmission Scheduling in Integrated mmWave Access and Terahertz Backhaul Networks}

\author{Lei~Wang,~\IEEEmembership{Graduate Student Member},
        Bo~Ai,~\IEEEmembership{Fellow,~IEEE},
        Yong~Niu,~\IEEEmembership{Senior Member,~IEEE},
        Haiyan~Jiang,
        Shiwen~Mao,~\IEEEmembership{Fellow,~IEEE},
        Zhangdui~Zhong,~\IEEEmembership{Fellow,~IEEE},
        and~Ning~Wang,~\IEEEmembership{Member,~IEEE}

\thanks{Copyright (c) 2015 IEEE. Personal use of this material is permitted. However, permission to use this material for any other purposes must be obtained from the IEEE by sending a request to pubs-permissions@ieee.org.}
\thanks{This work was supported by the Fundamental Research Funds for the Central Universities 2022YJS110; in part by the National Key Research and Development Program of China under Grant 2020YFB1806903; in part by the Fundamental Research Funds for the Central Universities under Grant 2022JBXT001 and Grant 2022JBQY004; in part by the National Key Research and Development Program of China under Grant 2021YFB2900301, Grant 2020YFB1806604, Grant 2021YFB3901302; in part by the National Natural Science Foundation of China under Grant 62221001, Grant 62231009, Grant U21A20445; in part by the Fundamental Research Funds for the Central Universities 2023JBMC03; in part by the Natural Science Foundation of Jiangsu Province Major Project under Grant BK20212002. (\emph{Corresponding authors: B. Ai, Y. Niu.})}
\thanks{L.~Wang is with the State Key Laboratory of Advanced Rail Autonomous Operation and the Beijing Engineering Research Center of High-Speed Railway Broadband Mobile Communications, Beijing Jiaotong University, Beijing 100044, China (e-mail: lleiwang@bjtu.edu.cn).}
\thanks{B.~Ai is with the State Key Laboratory of Advanced Rail Autonomous Operation, Beijing Jiaotong University, Beijing 100044, China, and also with the Research Center of Networks and Communications, Peng Cheng Laboratory, Shenzhen 518055, China, and also with the Henan Joint International Research Laboratory of Intelligent Networking and Data Analysis, Zhengzhou University, Zhengzhou 450001, China (e-mail: boai@bjtu.edu.cn).}
\thanks{Y.~Niu is with the State Key Laboratory of Advanced Rail Autonomous Operation, Beijing Jiaotong University, Beijing 100044, China, and also with the National Mobile Communications Research Laboratory, Southeast University, Nanjing 211189, China (e-mail: niuy11@163.com).}
\thanks{H.~Jiang and Z.~Zhong are with the State Key Laboratory of Advanced Rail Autonomous Operation, Beijing Jiaotong University, Beijing 100044, China (e-mails: 18120069@bjtu.edu.cn; zhdzhong@bjtu.edu.cn).}
\thanks{S.~Mao is with the Department of Electrical and Computer Engineering, Auburn University, Auburn, AL 36849-5201, USA (email: smao@ieee.org).}
\thanks{N.~Wang is with the School of Information Engineering, Zhengzhou University, Zhengzhou 450001, China (e-mail: ienwang@zzu.edu.cn).}
}
\maketitle

\begin{abstract}
Terahertz wireless backhaul is expected to meet the high-speed backhaul requirements of future ultra-dense networks using millimeter-wave (mmWave) base stations (BSs). In order to achieve higher network capacity with limited resources and meet the quality of service (QoS) requirements of more users in the integrated mmWave access and terahertz backhaul network, this paper formulates a problem of maximizing the number of users successfully served in both the access and backhaul links. Since the problem is a non-linear integer optimization problem, a minimum rate ratio user association and transmission scheduling algorithm is proposed to obtain a suboptimal solution. The proposed algorithm takes the minimum rate ratio as the user association criterion and schedules first the users with fewer backhaul transmission slots. In addition, the algorithm will update the number of access transmission slots allocated to users and the access scheduling results after the backhaul scheduling phase. Numerical results show that the proposed algorithm outperforms several benchmark algorithms in terms of the number of served users and system throughput, and it can cope with a large number of bursty user requests.
\end{abstract}


\section{Introduction}\label{S1}

Driven by the massive growth of mobile user equipments (UEs) and various new application scenarios, future communication systems are expected to be equipped with ultra-high speed access and backhaul. The large bandwidth in the millimeter wave (mmWave) band enables it to provide high data rates, ultra-reliability, and ultra-low latency, which are required by future wireless access networks \cite{mmwave}. Traditionally, these higher frequencies are not strong enough for outdoor broadband applications due to high propagation loss. They are also susceptible to blockage from buildings and human bodies. These problems made mmWave only reach out to a few kilometers. As a result, more mmWave base stations (BSs) equipped with large-scale antenna arrays need to be deployed to provide good coverage.

Ultra-dense network (UDN) improves network coverage and multiplies network throughput with the help of low-power small base stations (SBSs) \cite{udn}. However, it also brings about some inevitable challenges, such as increased cost. At present, most networks are backhauled through cables, among which, fiber-supported backhaul can provide a huge bandwidth and high capacity with low propagation loss. However, laying fiber for each BS in an UDN incurs a huge overhead, and it is even difficult to lay optical fiber in some areas \cite{fiber}. Compared with fiber-based backhaul, wireless backhaul is more flexible and convenient to deploy, which is more suitable for the future network architecture. The 3rd Generation Partnership Project (3GPP) has begun to evaluate solutions supporting wireless backhauls, and is committed to the standardization of integrated access and backhaul (IAB) technology \cite{iab}. The typical feature of IAB is that the access and backhaul links share wireless resources to improve spectrum utilization. The technology supports mmWave wireless backhaul to achieve dense cell deployment in areas with insufficient optical fiber, and helps to improve network capacity and coverage performance.

Unfortunately, high-capacity wireless access requires the support of a higher capacity wireless backhaul. MmWave frequency bands are insufficient to meet the high demands of 6G \cite{6g}. The available bandwidth of terahertz exceeds that of mmWave by an order of magnitude, and the terahertz bands can support extremely high backhaul data rates up to hundreds of gigabits per second or even terabits per second \cite{intro1}. Terahertz communications are regarded as a potential technology for 6G due to its ability to enable innovative applications in different scenarios. The IEEE 802.15.3d Task Group has carried out relevant research on terahertz communications\cite{intro2}. High-gain directional antennas, such as ultra-massive multiple-input multiple-output, are being designed to compensate for the high propagation loss of terahertz signals and extend the communication range \cite{intro3}.

Sara \emph{et al}. in \cite{sara} introduce the next-generation mobile heterogeneous network (HetNet) architecture, which is the coexistence of mmWave, terahertz, and traditional microwave. The authors envision a new communication paradigm that is completely different from the current one. A medium access protocol is designed to support switching between three coexisting frequency bands, and the selection of frequency bands is determined by the communication distance. The performance of the protocol is evaluated in a realistic scenario, where vehicles communicate with the infrastructure and traffic is backhauled to the data center. Therefore, one possible deployment scenario for next-generation networks is the integration of mmWave access and terahertz backhaul. The characteristics of the network put forward higher requirements for resource allocation. It is important and challenging to design an appropriate transmission scheduling scheme to avoid the waste of wireless resources. Moreover, user association and slot scheduling should be considered jointly to improve the network efficiency.

In this paper, we consider an IAB network architecture in which access links operate in mmWave frequency bands, while backhaul links use terahertz bands to provide a high-capacity backhaul. The network consists of an macro base station (MBS) and multiple SBSs. It is assumed that the UEs in the network communicate with the MBS through one of the SBSs, and the MBS is connected to the core network through optical fiber. To meet the quality of service (QoS) requirements of more users, we propose a joint user association and transmission scheduling algorithm. The main contributions of this paper are as follows.

\begin{itemize}
\item We investigate the joint user association and transmission scheduling problem of an integrated mmWave access and terahertz backhaul network. To meet the QoS requirements of more users in the network, a non-linear integer optimization problem is formulated to maximize the number of served users, where a served user is defined as a user whose throughput on both the access link and the backhaul link meets the QoS requirements. The optimization problem is constrained by the limited time slot resources and the half-duplex mode of BSs.

\item A minimum rate ratio user association and IAB transmission scheduling algorithm is proposed to obtain a suboptimal solution of the problem. The SBS makes the pre-decision of access transmission scheduling following the minimum rate ratio user association policy. In the backhaul scheduling phase, considering the finite time resources and the QoS requirements of users, the links that require fewer backhaul transmission slots are preferentially scheduled. In particular, the algorithm updates the number of access slots after backhaul transmission scheduling to further improve network capacity.

\item We examine the effect of system parameters on the performance of the proposed algorithm. Simulation results show that the proposed algorithm achieves a better performance than several baseline schemes in terms of the number of served users and system throughput, and is able to handle a large number of bursty user requests.
\end{itemize}

The remainder of this paper is organized as follows. Section~\ref{S1-1} summarizes the related research on integrated access and backhaul. Section~\ref{S2} presents the system model and formulates a problem to maximize the number of served users, which is a non-linear integer optimization problem. A joint user association and transmission scheduling algorithm is proposed in Section~\ref{S3}. Section~\ref{S4} compares the proposed algorithm with several benchmark schemes to demonstrate its effectiveness. Finally, Section~\ref{S5} concludes this paper. The main mathematical notations of this paper are given in Table \ref{notation}.

\section{Related Work}\label{S1-1}
3GPP envisions an IAB architecture for 5G cellular networks to enable flexible network deployment. In an IAB network, the user association policy %is also closely related to the resource utilization of the network, which
greatly affects the achievable data rate of the user and the throughput of the entire network. The authors of \cite{ua1} characterized the correlation between user transmission rate and user behavior parameters, and made user association decisions based on the network state and user behavior parameters. The problem of maximizing network throughput was formulated as a stochastic optimization problem. Ref. \cite{ua3} focused on HetNets with access links operating in multiple mmWave frequency bands, and a Markov approximation framework was used to solve the joint user association and power allocation problems. In view of the interaction between user association and user scheduling, Ge \emph{et al}. in \cite{ua4} proposed an algorithm based on the alternating direction method of multipliers for network load balancing. In \cite{ua2}, a deep reinforcement learning (DRL)-based algorithm was introduced to solve the joint user association and channel allocation problem in HetNets to reduce the communication overhead of the network.

Terahertz band is expected to provide Tbps-level data rates for cellular and backhaul communications due to its ultra-large bandwidth \cite{thzb1}. Some studies have been devoted to optimizing the user association and scheduling mechanisms in terahertz communication systems.
In indoor terahertz communication networks, user access may fail due to blockage by the user itself, surrounding mobile users and walls, or benefit from the 3D directional antennas that access points and UEs are equipped with. The coverage performance of the terahertz downlink in indoor environment was derived in \cite{thzloss}. To serve far more users than radio frequency (RF) chains in a terahertz communication system, a cluster-based multi-carrier hybrid beamforming scheme was designed in \cite{thzb2}, in which users were scheduled to different clusters based on their location information. The scheme maximized the throughput of the system through a dynamic connection structure between RF chains and antenna subarrays. The signal-to-noise ratio (SNR) and ergodic capacity of a terahertz-RF link was analyzed in \cite{thz3}, where the UE was associated to the access point over RF and the UE traffic was backhauled to the core network over terahertz. In \cite{thz1}, a greedy shrinking algorithm was developed for slots, bandwidth scheduling and power allocation in a terahertz self-backhaul network to maximize network throughput.

Resource allocation is also a hot topic in IAB research. In \cite{ra2}, the expression of rate coverage for dynamic bandwidth partitioning between access and backhaul links in IAB networks was derived. The results demonstrated that it is undesirable to increase the data rate of users by increasing the density of SBS. Spectrum resources were also allocated through DRL approaches in \cite{ra3}. This framework was suitable for different IAB network structures and enabled real-time spectrum allocation. Tan \emph{et al}. in \cite{ra4} focued on user's quality of experience, and proposed to temporarily activate a nearby IAB node or switch to a new IAB node for UEs with low QoS. Ref. \cite{ra5} constructed the topology of the network through the measurement values from IAB nodes, and then assigned different priorities to each link for resource allocation through the maximum weighted matching algorithm.

However, most algorithms were designed for low frequency or mmWave communications; they may not be suitable for scheduling in integrated mmWave access and terahertz backhaul networks. In addition, in the research on the resource allocation in IAB networks, few prior works took different QoS requirements of users into consideration, and the same QoS requirements cannot reflect the real situation of the network.

\begin{figure}[!t]
\begin{center}
\includegraphics*[width=0.95\columnwidth,height=2.2in]{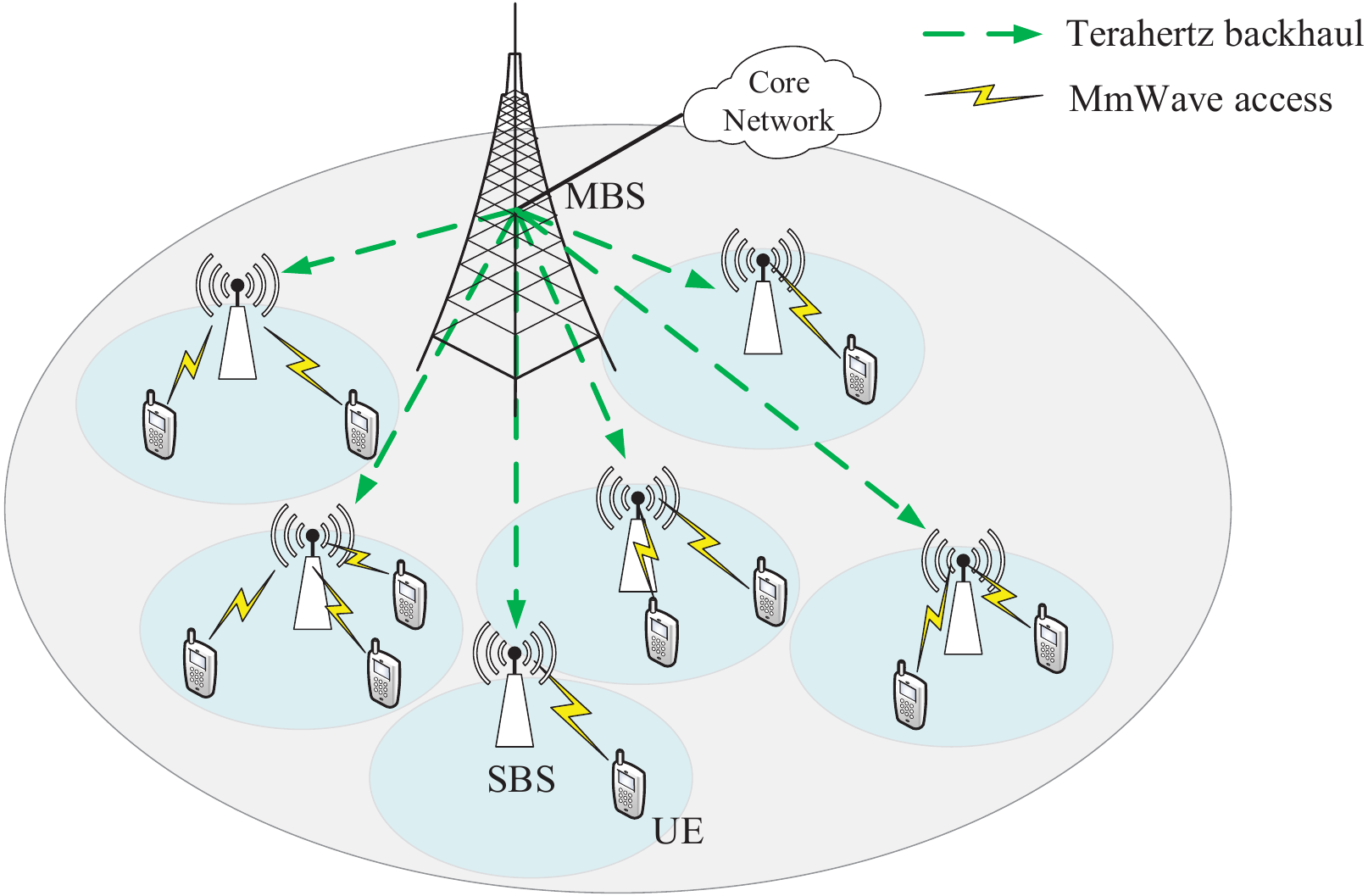}
\end{center}
\caption{Integrated mmWave access and terahertz backhaul network.}
\label{network}
\end{figure}

\begin{figure}[!t]
\begin{center}
\includegraphics*[width=0.95\columnwidth,height=1.2in]{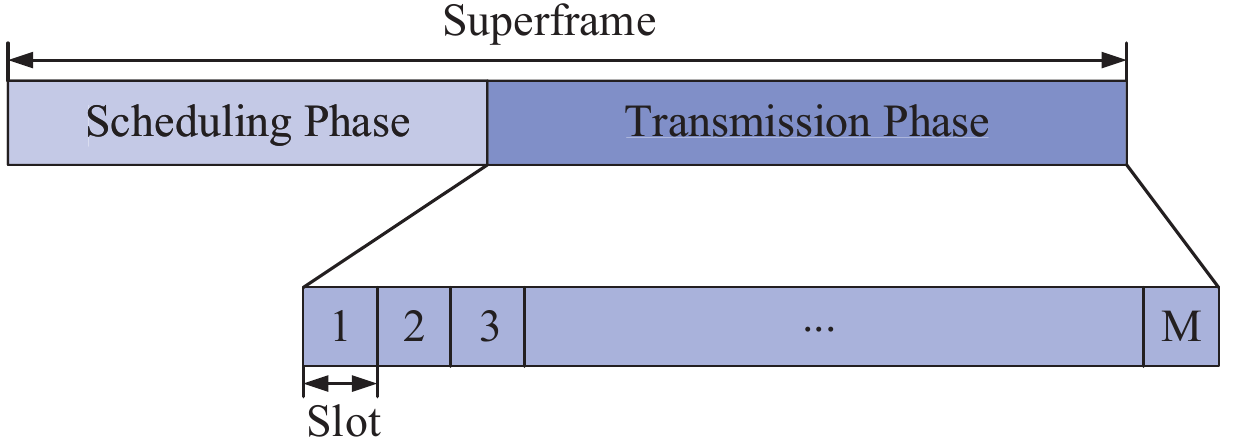}
\end{center}
\caption{The superframe structure of the backhaul adopted in this paper.}
\label{frame}
\end{figure}

\section{System Model And Problem Formulation}\label{S2}

This section presents the system model and formulates the transmission scheduling problem for an integrated mmWave access and terahertz backhaul network.

\subsection{System Model}\label{S2-1}

We consider a cellular network composed of one MBS, $L$ SBSs, and $K$ uniformly distributed UEs in a two dimensional space $\mathbb{R}^2$, as shown in Fig. \ref{network}. The MBS serves UEs within its coverage area over wireless links and is connected to the core network through wired fiber. Each SBS acts as an IAB node, then an access link refers to a link between a UE and an SBS. A backhaul link refers to a link between an SBS and the MBS. It is assumed that all UEs are connected to the MBS through a two-hop link, that is, the traffic of UEs is first routed to an SBS through the access link, and then forwarded to the MBS through the backhaul link. Furthermore, access traffic and backhaul traffic are carried over different frequency bands. Access links are assumed to operate in the mmWave frequency bands, and backhaul links are assumed to operate in the terahertz frequency bands to support ultra-high data rate transmission. As a result, there is no interference between the simultaneous transmissions of access links and backhaul links. It is assumed that all BSs and UEs are equipped with antenna arrays to compensate for the large path loss at these high frequencies.

In this system, time is partitioned into non-overlapping superframes. Each superframe is divided into a scheduling phase and a transmission phase, and each phase consists of multiple slots, which is illustrated in Fig. \ref{frame}. Specifically, it is assumed that the scheduling phase of an access frame and a backhaul frame have the same duration, denoted by $t_s$, and the transmission phases of the two consist of $N$ and $M$ equal length slots, respectively. The duration of each slot is denoted by $\Delta$, then the duration of an access superframe is ${t_A} = {t_s} + N  \Delta $, and the duration of a backhaul superframe is ${t_B} = {t_s} + M \Delta $.

\linespread{1.5}
\begin{table}[!t]
\begin{center}
\caption{Summary of Main Mathematical Symbols}
\begin{tabular}{ll}
\toprule
{\em Symbol} & {\em Description}\\
\midrule
$t_s$ & Duration of the scheduling phase \\
$\Delta$ & Duration of each slot \\
$N$ & \makecell[l]{Number of slots in the transmission phase of \\ an access frame} \\
$M$ & \makecell[l]{Number of slots in the transmission phase of \\ a backhaul frame} \\
$P_t^A$, $P_t^B$ & Transmit power of access links and backhaul links  \\
$G_{tx}$, $G_{rx}$ & Antenna gains of transmitter and receiver  \\
$P_{{b_l},{u_k}}^A$ & Received power at $u_k$ from $b_l$  \\
$R_{{b_l},{u_k}}^A$ & Downlink data rate of $u_k$ from $b_l$  \\
$R_{{b_0},{b_{lk}}}^B$ & Data rate of ${b_l}$ associated with $u_k$ \\
$x_{{b_l},{u_k}} $ & Binary user-association variable \\
$a_{_{{b_l},{u_k}}}^A $, $a_{_{{b_0},{b_{lk}}}}^B$ & Binary scheduling variables \\
$T_{{b_l},{u_k}}^A$ & Number of slots allocated to the access link \\
$T_{{b_0},{b_{lk}}}^B$ & Number of slots allocated to the backhaul link \\
$C_{{u_k}}^A$, $C_{{u_k}}^B$ & Throughput on the access and backhaul link \\
$I_{u_k}$ & Binary served-user variable \\
$t_{{b_l},{u_k}}^A$ & Number of slots required for the access link \\
$t_{{b_0},{b_{lk}}}^B$ & Number of slots required for the backhaul link \\
\bottomrule
\end{tabular}
\label{notation}
\end{center}
\end{table}

Let $\mathcal{B}$ denote the set of SBSs, as $ \mathcal{B}= \left\{ {{b_1},{b_2}, \ldots ,{b_L}} \right\}$, where $b_l,l \in \mathcal{L}$ is the $l$th SBS and $\mathcal{L} = \left\{ {1,2, \ldots ,L} \right\}$. Let $\mathcal{U}$ denote the set of UEs, as $ \mathcal{U}= \left\{ {{u_1},{u_2}, \ldots ,{u_K}} \right\}$, where $u_k,k \in \mathcal{K}$ is the $k$th UE and $\mathcal{K} = \left\{ {1,2, \ldots ,K} \right\}$. The MBS is denoted by $b_0$.

For an access link with $b_l$ as the transmitter (TX) and $u_k$ as the receiver (RX), we denote the antenna gain of $b_l$ by ${G_{tx}}\left( {{\theta _{lk}}} \right)$, and the antenna gain of $u_k$ by ${G_{rx}}\left( {{\theta _{lk}}} \right)$, where ${\theta _{lk}}$ is the off-axis angle. Accordingly, the received power at $u_k$ from $b_l$, denoted by $P_{{b_l},{u_k}}^A$, is given by
\begin{equation}
P_{{b_l},{u_k}}^A = \beta P_t^A{G_{tx}}\left( {{\theta _{lk}}} \right){G_{rx}}\left( {{\theta _{lk}}} \right)d_{lk}^{ - \alpha },
\end{equation}
where $\beta \propto {\left( {{\lambda  \mathord{\left/{\vphantom {\lambda  {4\pi }}} \right. \kern-\nulldelimiterspace} {4\pi }}} \right)^2}$ is a constant coefficient, $\lambda$ is the wavelength, $P_t^A$ is the transmit power of access links, $d_{lk}$ is the distance between $u_k$ and $b_l$, and $\alpha$ is the path loss exponent. Since the directional transmission reduces the inter-link interference, the downlink data rate of $u_k$ from $b_l$, denoted by $R_{{b_l},{u_k}}^A$, can be achieved as
\begin{equation}
R_{{b_l},{u_k}}^A = \eta {W_A}{\log _2}\left( {1 + \frac{{\beta P_t^A{G_{tx}}\left( {{\theta _{lk}}} \right){G_{rx}}\left( {{\theta _{lk}}} \right)d_{lk}^{ - \alpha }}}{{{N_0}{W_A}}}} \right),
\end{equation}
where $\eta  \in \left( {0,1} \right)$ is the efficiency of the TX, $W_A$ is the bandwidth of access links, and $N_0$ is the noise power spectral density.

Similarly, we denote $P_t^B$ as the transmit power of backhaul links, and $W_B$ as the bandwidth of backhaul links. Then the data rate of the backhaul link from ${b_0}$ to ${b_l}$ associated with $u_k$, denoted by $R_{{b_0},{b_{lk}}}^B$, is given by
\begin{equation}
R_{{b_0},{b_{lk}}}^B = \eta {W_B}{\log _2}\left( {1 + \frac{{ P_t^B{G_{tx}}\left( {{\phi  _{0l}}} \right){G_{rx}}\left( {{\phi _{0l}}} \right)\rho \left( {{b_0},{b_l}} \right)}}{{{N_0}{W_B}}}} \right),
\end{equation}
where ${G_{tx}}\left( {{\phi _{0l}}} \right)$ and ${G_{rx}}\left( {{\phi _{0l}}} \right)$ are the antenna gains of TX and RX, respectively, $\phi _{0l}$ is the off-axis angle, and $\rho \left( {{b_0},{b_l}} \right)$ denotes the path loss from ${b_0}$ to ${b_l}$.

\subsection{Problem Formulation}\label{S2-2}

In the considered IAB-enabled networks, SBSs act as relays between UEs and the MBS. After a UE initiates a communication request, it first associates with an SBS before establishing an access link. Multiple UEs may connect to different SBSs according to the user association policy. For the $u_k, \forall k \in \mathcal{K}$, the binary user-association vector is defined as ${{\bf{x}}_{b,{u_k}}}= \left( {{x_{{b_1},{u_k}}},{x_{{b_2},{u_k}}}, \ldots ,{x_{{b_L},{u_k}}}} \right)$, where
\begin{equation}
{x_{{b_l},{u_k}} \left( t \right)} = \left\{ \begin{array}{l}
1,\ \mbox{if $u_k$ associates with $b_l$}\\
0,\ \mbox{otherwise}.
\end{array} \right.\label{assocaition}
\end{equation}

Assume that each UE can only associate with one SBS at time $t$, that is,
\begin{equation}
\sum\limits_{l = 1}^L {{x_{{b_l},{u_k}}}\left( t \right)}  \le 1,\forall k \in \mathcal{K}.\label{ass cons}
\end{equation}

After user association, the SBS and the MBS will schedule the access link and the backhaul link respectively. A binary scheduling variable is given by
\begin{equation}
a_{_{{b_l},{u_k}}}^A\left( t \right) = \left\{ \begin{array}{l}
1,\ \mbox{if the access link $(b_l,u_k)$ is scheduled}\\
0,\ \mbox{otherwise}.
\end{array} \right.\label{a_access}
\end{equation}

Let $a_{_{{b_0},{b_{lk}}}}^B$ denote whether data from $u_k$ associated with the SBS $b_l$ is scheduled in the backhaul scheduling phase,
\begin{equation}
a_{_{{b_0},{b_{lk}}}}^B\left( t \right) = \left\{ \begin{array}{l}
1,\ \mbox{if the data from $u_k$ is scheduled}\\
0,\ \mbox{otherwise}.
\end{array} \right.\label{a_backhaul}
\end{equation}

We assume that the MBS and SBSs operate under half-duplex constraints. If two different UEs $u_i$ and $u_j$ are both associated with the same SBS $b_l$, the potential access links $(b_l,u_i)$ and $(b_l,u_j)$ cannot be scheduled to transmit at the same time, i.e.,
\begin{equation}
a_{_{{b_l},{u_i}}}^A\left( t \right) + a_{_{{b_l},{u_j}}}^A\left( t \right) \le 1,\forall l \in L,i \neq j.\label{acc1}
\end{equation}

Similarly, backhaul links can only be transmitted sequentially,
\begin{equation}
a_{_{{b_0},{b_{lk}}}}^B\left( t \right) + a_{_{{b_0},{b_{nm}}}}^B\left( t \right) \le 1,\forall l \neq n,k \neq m.\label{back}
\end{equation}

We denote the number of slots allocated to the access link $(b_l,u_i)$ and the backhaul link $(b_0,b_{lk})$ by $T_{{b_l},{u_k}}^A$ and $T_{{b_0},{b_{lk}}}^B$, respectively. Since the number of transmission slots is limited and each slot can only be occupied by one link, we have
\begin{equation}
\sum\limits_{k = 1}^K {T_{{b_l},{u_k}}^A}  \le N,\forall l \in \mathcal{L}.\label{N}
\end{equation}
\begin{equation}
\sum\limits_{l = 1}^L {\sum\limits_{k = 1}^K {T_{{b_0},{b_{lk}}}^B} }  \le M.\label{M}
\end{equation}

Consequently, the throughput achieved by UE $u_k$ on the access link, denoted by $C_{{u_k}}^A$, can be expressed as
\begin{equation}
C_{{u_k}}^A = \frac{{R_{{b_l},{u_k}}^A  T_{{b_l},{u_k}}^A  \Delta }}{{{t_s} + N  \Delta }}.\label{C}
\end{equation}

The throughput achieved by UE $u_k$ on the backhaul link, denote by $C_{{u_k}}^B$, can be expressed as
\begin{equation}
C_{{u_k}}^B = \frac{{R_{{b_0},{b_l}_k}^B  T_{{b_0},{b_l}_k}^B  \Delta }}{{{t_s} + M  \Delta }}.
\end{equation}

In an IAB network, a user can be successfully served only if the achievable throughput of the access link and the backhaul link meet the user's minimum QoS requirements ${\Omega _{{u_k}}}$, that is, $C_{{u_k}}^A \ge {\Omega _{{u_k}}}$ and $C_{{u_k}}^B \ge {\Omega _{{u_k}}}$. The user who meets the above conditions is called a served user. A binary variable $I_{u_k}$ is defined to indicate whether $u_k$ is a served user, which is given by
\begin{equation}
{I_{u_k}} = \left\{ \begin{array}{l}
1,\ \mbox{if $C_{{u_k}}^A \ge {\Omega _{{u_k}}}$\ $\&$\ $C_{{u_k}}^B \ge {\Omega _{{u_k}}}$}\\
0,\ \mbox{otherwise}.
\end{array} \right.
\end{equation}

From the perspective of improving user experience, the optimization problem is formulated as
\begin{subequations}
\begin{align}
\max \quad&\sum\limits_{k \in \mathcal{K}} {{I_{u_k}}} \\
\rm s.t.\quad&{x_{{b_l},{u_k}}} \in \left\{ {0,1} \right\},\forall l,k,\label{b}\\
&\sum\nolimits_{l = 1}^L {{x_{{b_l},{u_k}}}\left( t \right)}  \le 1,\forall k,\label{c}\\
&a_{{b_l},{u_k}}^A \left( t \right)\in \left\{ {0,1} \right\},\forall l,k,t,\label{d}\\
&a_{{b_0},{b_{lk}}}^B\left( t \right) \in \left\{ {0,1} \right\},\forall l,k,t,\label{e}\\
&a_{_{{b_l},{u_i}}}^A\left( t \right) + a_{_{{b_l},{u_j}}}^A\left( t \right) \le 1,\forall l ,i \neq j,\label{f}\\
&a_{_{{b_0},{b_{lk}}}}^B\left( t \right) + a_{_{{b_0},{b_{nm}}}}^B\left( t \right) \le 1,\forall l \neq n, k \neq m,\label{g}\\
&\sum\nolimits_{k = 1}^K {T_{{b_l},{u_k}}^A}  \le N,\forall l,\label{h}\\
&\sum\nolimits_{l = 1}^L {\sum\nolimits_{k = 1}^K {T_{{b_0},{b_{lk}}}^B} }  \le M,\label{i}
\end{align}
\end{subequations}
where constraints \eqref{b}, \eqref{d}, and \eqref{e} define the user association variable and two transmission scheduling variables, respectively. Constraint \eqref{c} ensures that the UE can only associate with one SBS at any time. Constraint \eqref{f} and \eqref{g} indicate that the MBS and SBSs operate in the half-duplex mode. Constraint \eqref{h} and \eqref{i} guarantee that the number of scheduled slots does not exceed the number of transmission slots per frame. Since the problem is a non-linear integer optimization problem, it is challenging to design the optimal joint user association and transmission scheduling scheme. In the next section, we will propose a heuristic algorithm to obtain a suboptimal solution to this problem.

\section{Joint User Association and Transmission Scheduling Algorithm}\label{S3}

The purpose of the algorithm design is to maximize the number of served users. Since the number of access and backhaul transmission slots is limited, more UEs can be scheduled if the UE occupies fewer transmission slots of the access frame and the backhaul frame. The initial user association results will affect the scheduling of access links and backhaul links. If UE $u_k$ is associated with SBS $b_l$, the number of transmission slots required for access link $(b_l, u_k)$, denoted by $t_{{b_l},{u_k}}^A$, is calculated by
\begin{equation}
t_{{b_l},{u_k}}^A = \frac{{{\Omega _{{u_k}}}\left( {{t_s} + N  \Delta } \right)}}{{R_{{b_l},{u_k}}^A  \Delta }}.\label{t1}
\end{equation}

The number of transmission slots required for the backhaul link $(b_0, b_{lk})$, denoted by $t_{{b_0},{b_{lk}}}^B$, is computed by
\begin{equation}
t_{{b_0},{b_{lk}}}^B = \frac{{{\Omega _{{u_k}}}\left( {{t_s} + M  \Delta } \right)}}{{R_{{b_0},{b_{lk}}}^B  \Delta }}.\label{t2}
\end{equation}

From \eqref{t1} and \eqref{t2}, it can be found that if UE $u_k$ is associated with the SBS with the smallest ${{\Omega _{{u_k}}}}/{R_{{b_l},{u_k}}^A}$ and ${{\Omega _{{u_k}}}}/{R_{{b_0},{b_{lk}}}^B}$, the number of transmission slots required for the access and backhaul link will be the smallest. In practice, it would be difficult to find an SBS that satisfies both of the above conditions. Considering that the communication request is initiated by the user, we formulate a user association policy based on ${{\Omega _{{u_k}}}}/{R_{{b_l},{u_k}}^A}$, which is defined as the rate ratio (QoS/Rate) and abbreviated as QR in the remainder of this paper.

\linespread{1.2}
\begin{algorithm}[t!]
%\setstretch{1.1}
	\caption{Minimum Rate Ratio User Association and Transmission Scheduling Algorithm}
	\label{alg1}
	\begin{algorithmic}[1]
	\REQUIRE
	locations of the MBS, $L$ SBSs and $K$ UEs; QoS requirements for each user
	\ENSURE
	user association and transmission scheduling results
		\WHILE {$\left| \mathcal{B} \right| \ne \emptyset \ \&\ \left| \mathcal{U} \right| \ne \emptyset$}
		\FOR {$l=1$ to $L$}
		\STATE Calculate the rate ratio between the $b_l$ and all unassociated UEs and select the UE $u_k$ with the smallest value as the candidate associated UE;
		\STATE $\mathcal{U}_{b_l}=\mathcal{U}_{b_l}+\left\{ u_k \right\}$;
		\STATE Calculate $T_{{b_l},{u_k}}^A$ to be allocated to $u_k$;
		\IF {$\sum\limits_{u \in {\mathcal{U}_{b_l}}} {T_{{b_l},{u_k}}^A}  \le N$}
		\STATE $x_{{b_l},{u_k}}=1$,\ $\mathcal{U}=\mathcal{U}-\left\{ u_k \right\}$;
		\ELSE
		\STATE $\mathcal{U}_{b_l}=\mathcal{U}_{b_l}-\left\{ u_k \right\}$,\ $\mathcal{B}=\mathcal{B}-\left\{ b_l \right\}$;
		\ENDIF
		\ENDFOR
		\ENDWHILE
		\FOR {$l=1$ to $L$}
		\FOR {$u_i \in {\mathcal{U}_{b_l}}$}
		\STATE Calculate $T_{{b_0},{b_{li}}}^B$ to be allocated to $u_i$;
		\ENDFOR
		\ENDFOR
		\STATE Calculate the total number of backhaul transmission slots for all associated UEs $T = \sum\nolimits_{l = 1}^L {\sum\nolimits_{{u_i} \in {\mathcal{U}_{b_l}}} {T_{{b_0},{b_{li}}}^B} }  $;
		\WHILE {$T>M$}
		\STATE Find the UE ${u_j}$ with the largest $T_{{b_0},{b_{mj}}}^B$;
		\STATE $\mathcal{U}_{b_m}=\mathcal{U}_{b_m}-\left\{ u_j \right\}$,\ $T=T-T_{{b_0},{b_{mj}}}^B$,\ $x_{{b_m},{u_j}}=0$;
		\ENDWHILE
		\FOR {$l=1$ to $L$}
		\FOR {$u_i \in {\mathcal{U}_{b_l}}$} \vspace{0.7ex}
		\STATE Update $\tilde T_{{b_l},{u_i}}^A$; \vspace{0.7ex}
		\STATE Calculate $C_{{u_i}}^A$, $C_{{u_i}}^B$;
		\ENDFOR
		\ENDFOR			
	\end{algorithmic}
\end{algorithm}

According to \eqref{C} and \eqref{t1}, if the access transmission slots allocated to the UE are more than the required access transmission slots, the throughput achieved by the UE on the access link will be greater than the required minimum throughput. As a result, the number of access transmission slots initially allocated to each UE can be obtained by rounding up the required number of access transmission slots. Then we have
\begin{equation}
T_{{b_l},{u_k}}^A = \left\lceil {\frac{{{\Omega _{{u_k}}}\left( {{t_s} + N  \Delta } \right)}}{{R_{{b_l},{u_k}}^A  \Delta }}} \right\rceil,\label{rt1}
\end{equation}
where $\left\lceil x \right\rceil  = \min \left\{ {n \in \mathbb{Z}\left| {x \le n} \right.} \right\}$.

The allocation principle of backhaul transmission slots is based on the QoS requirements of users. As the achievable throughput of an SBS depends on the minimum throughput of the access link and the backhaul link, the throughput of the backhaul link should also meet the QoS requirements of users. Then the number of backhaul transmission slots allocated to each backhaul link is given by
\begin{equation}
T_{{b_0},{b_{lk}}}^B = \left\lceil {\frac{{{\Omega _{{u_k}}}\left( {{t_s} + M  \Delta } \right)}}{{R_{{b_0},{b_{lk}}}^B  \Delta }}} \right\rceil.\label{rt2}
\end{equation}

A minimum rate ratio user association and transmission scheduling algorithm is proposed, as shown in Algorithm \ref{alg1}. The main steps of the algorithm are as follows.

\emph{1) User association and access link scheduling pre-decision}: User association is based on the minimum rate ratio. An SBS is associated with the UE with the smallest rate ratio among all unassociated UEs, as shown in line 3. This process will be repeated until all SBSs complete the user association. Note that only associated users satisfying \eqref{N} can be pre-scheduled, and this judgment is done by line 6. Through lines 1 to 12, the user association results and the scheduling pre-decision of the access links can be obtained, as well as the number of slots allocated to each access link according to \eqref{rt1}.

\emph{2) Backhaul link transmission scheduling}: This step first assumes that the backhaul traffic of all associated UEs can be scheduled and calculates the backhaul transmission slots allocated to each UE according to \eqref{rt2}, as in lines 13 to 17. Due to limited backhaul transmission slots, the QoS requirements of all associated UEs cannot be fully met. From lines 19 to 22, if the total number of backhaul transmission slots allocated to all associated UEs is greater than $M$, the algorithm cancels the scheduling of the UEs occupying the most slots until \eqref{M} is satisfied, that is, the UE who requires fewer slots are preferentially scheduled.

\emph{3) User association and transmission scheduling update}: The access link scheduling for UEs whose backhaul link is not scheduled will be cancelled. The algorithm then releases the access transmission slots originally allocated to the unscheduled UEs, and updates the number of access slots of scheduled UEs, as shown in lines 23 to 28.

The proposed algorithm makes full use of slot resources and further improves the throughput of the system by updating the number of access slots. The update method is as follows.

Assume that the set of UEs associated with SBS $b_l$ is $\mathcal{U}_{b_l}$, the updated number of access slots allocated to $u_k$ is
\begin{equation}
\tilde T_{{b_l},{u_k}}^A = \left\lfloor {\frac{N {T_{{b_l},{u_k}}^A}}{{\sum\limits_{{u_k} \in {\mathcal{U}_{b_l}}} {T_{{b_l},{u_k}}^A} }} } \right\rfloor,\label{updata}
\end{equation}
where $\left\lfloor x \right\rfloor  = \max \left\{ {n \in \mathbb{Z} \left| {n \le x} \right.} \right\}$.

\section{Performance Evaluation}\label{S4}

In this section, we evaluate the performance of the proposed minimum rate ratio user association and transmission scheduling algorithm in terms of the number of served users and system throughput. Specifically, we evaluate the impact of system parameters including the number of UEs, the number of transmission slots in an access frame and a backhaul frame, and the access and backhaul transmit power on the performance of the proposed algorithm. The key system parameters are summarized in Table \ref{table1}. The value of the backhaul bandwidth is determined by the operating frequency and transmission window \cite{thzband2}.

\linespread{1.2}
\begin{table}[!t]
\begin{center}
\caption{Simulation Parameters}
\begin{tabular}{lcc}
\toprule
{\em Parameter} & {\em Symbol}  & {\em Value} \\
\midrule
%Transmit power & $P_t $ & 1000mW  \\
Transceiver transmission efficiency factor & $\eta $ & 0.9  \\
Operating frequency at the access link & $f_A$ & 62-64 GHz \cite{mmband} \\
Access bandwidth & $W_A$ & 2 GHz \\
Operating frequency at the backhaul link & $f_B$ & 300-320 GHz \cite{thzband}\\
Backhaul bandwidth &  $W_B$  & 20 GHz \\
Noise power spectral density & $N_0$ & -134\ dBm/MHz \\
Path loss exponent &  $\alpha $   &  2 \\
Slot duration & $\Delta $ & $18\ \mu s$ \\
Scheduling phase duration&  $t_s$ & $850\ \mu s$  \\
Half-power beamwidth & $\theta_{-3dB}$ & $30^ \circ$ \\
\bottomrule
\end{tabular}
\label{table1}
\end{center}
\end{table}

\subsection{Simulation Setup}\label{S4-1}

Assume that there are 8 SBSs and no more than 500 UEs uniformly distributed in a 100 $ \times $ 100 $m^2$ area. Considering that the user experience rate will reach 1 Gbps or even higher in the near future \cite{gbps}, it is assumed that the QoS requirements of each user are uniformly distributed between $\left[ {2,5} \right]$ Gbps. We take the average value of 50 executions of the algorithm as the final result. The topology of the network, the locations of the SBSs, the locations of UEs, and the QoS requirements of users are varied in each simulation.

The realistic directional antenna model is adopted for mmWave access links \cite{antenna}. The directional antenna gain, denoted by $G(\theta)$, is given by
\begin{equation}
G(\theta)=\left\{
\begin{array}{ll}
G_0-3.01  {\left(\dfrac{2\theta}{\theta_{-3dB}}\right)}^{2},& {0^\circ\leq\theta\leq{\theta_{ml}}/2}, \\ %\vspace{1.3ex}
G_{sl},& {{\theta_{ml}}/2\leq\theta\leq180^\circ},
\end{array}
\right.
\end{equation}
where $\theta$ is the off-axis angle, $\theta_{-3dB}$ is the angle of the half-power beamwidth, and $\theta_{ml}$ is the main lobe width and satisfies ${\theta _{ml}} = 2.6 \cdot {\theta _{ - 3dB}}$. The maximum antenna gain $G_0$ is computed as ${G_0} = 10\lg {\left( {{{1.6162} \mathord{\left/{\vphantom {{1.6162} {\sin \left( {{{{\theta _{ - 3dB}}} \mathord{\left/{\vphantom {{{\theta _{ - 3dB}}} 2}} \right. \kern-\nulldelimiterspace} 2}} \right)}}} \right. \kern-\nulldelimiterspace} {\sin \left( {{{{\theta _{ - 3dB}}} \mathord{\left/{\vphantom {{{\theta _{ - 3dB}}} 2}} \right.\kern-\nulldelimiterspace} 2}} \right)}}} \right)^2}$. The side lobe gain $G_{sl}$ can be expressed as ${G_{sl}} =  - 0.4111 \cdot \ln \left( {{\theta _{ - 3dB}}} \right) - 10.579$.

The ITU-R Recommendation F.699-7 narrow beam antenna model is adopted for terahertz backhaul links \cite{antenna2}. The antenna gain relative to an isotropic antenna, denoted by $G\left( \phi  \right)$, is given by
\begin{equation}
G\left( \phi  \right) = \left\{ \begin{array}{llll}
{G_{\max }} - 2.5 \times {10^{ - 3{{\left( {{D}/{\lambda }\phi } \right)}^2}}},&{0^ \circ } < \phi  < {\phi _m}, \vspace{0.7ex}\\
{G_1},&{\phi _m} \le \phi  < {\phi _r}, \vspace{0.7ex}\\
32 - 25\log \phi ,&{\phi _r} \le \phi  < {48^ \circ }, \vspace{0.7ex}\\
 - 10,&{48^ \circ } \le \phi < {180^ \circ },
\end{array} \right.
\end{equation}
where $\phi$ is the off-axis angle, $G_{\max }$ is the maximum antenna gain, $D$ is antenna diameter and $\lambda$ is wavelength. $G_1$ is the gain of the first side-lobe and is obtained by ${G_1} = 2 + 15\lg \left( {{D \mathord{\left/{\vphantom {D \lambda }} \right.\kern-\nulldelimiterspace} \lambda }} \right)$. In the simulation, Cassegrain antennas are deployed on the MBS and SBSs, with ${G_{\max }} = 47\ dBi$ and ${D \mathord{\left/{\vphantom {D \lambda }} \right.\kern-\nulldelimiterspace} \lambda } = 152$. Parameters ${\phi _m}$ and $\phi _r$ can be calculated by ${\phi _m} = {{20\lambda } \mathord{\left/{\vphantom {{20\lambda } D}} \right.\kern-\nulldelimiterspace} D}\sqrt {{G_{\max }} - {G_1}} $ and ${\phi _r} = 15.85{\left( {{D \mathord{\left/{\vphantom {D \lambda }} \right.\kern-\nulldelimiterspace} \lambda }} \right)^{ - 0.6}}$ respectively.

Since the terahertz backhaul link suffers from high molecular absorption loss, the path loss in the terahertz band is defined as the product of the spreading loss and the molecular absorption loss, which is given by  \cite{thzloss,model}
\begin{equation}
\rho\left(b_0, b_l\right)=\left(\frac{c}{4 \pi f_B d_{0 l}}\right)^2 e^{-k(f_B) d_{0 l}},\label{loss}
\end{equation}
where $c$ is the speed of light, $f_B$ is the operating frequency at the backhaul link, $d_{0l}$ is the distance between $b_0$ and $b_l$, and $k(f_B)$ is the medium absorption coefficient. In \eqref{loss}, $\left(c / 4 \pi f_B d_{0 l}\right)^2$ refers to the spreading loss, and $e^{-k(f_B) d_{0 l}}$ stands for the molecular absorption loss, that is, the energy attenuation of electromagnetic waves caused by molecular absorption. Specifically, at pressure $p$ and temperature $T$, $k(f_B)$ can be further computed by $k(f_B)=\frac{p}{p_{S T P}} \frac{T_{S T P}}{T} \sum_{i, g} Q^{i, g} \sigma^{i, g}(f_B)$, where $p_{S T P}$ and $T_{S T P}$ are the standard pressure and temperature, respectively, $Q^{i, g}$ is the molecular volumetric density of the isotopologue $i$ of gas $g$, and $\sigma^{i, g}(f_B)$ is the absorption cross section for the isotopologue $i$ of gas $g$ at the frequency $f_B$. Parameters $Q^{i, g}$ and $\sigma^{i, g}(f_B)$ can be obtained from the HITRAN (high resolution transmission molecular absorption) database \cite{rothman2009hitran}.

\begin{figure}[!t]
    \begin{center}
        \includegraphics[width=0.9\columnwidth,height=2.5in]{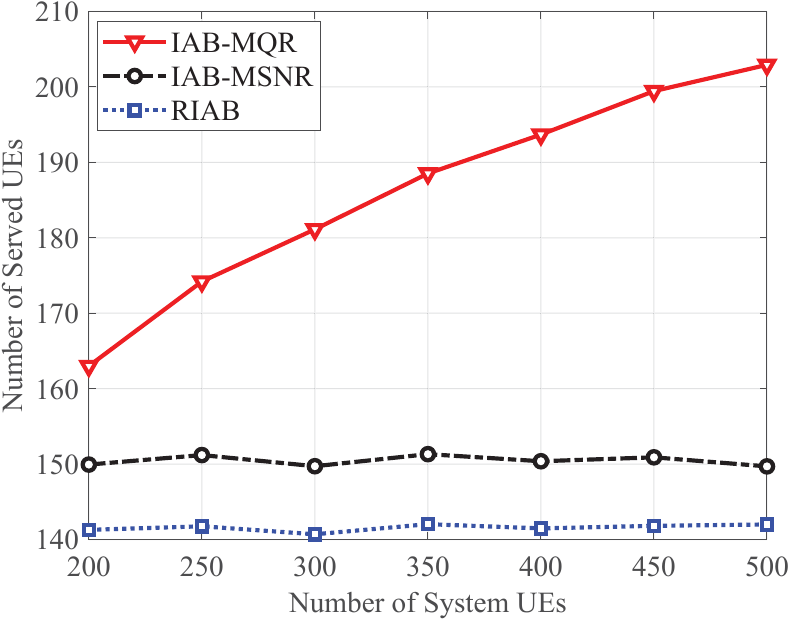}
        \caption{The number of served UEs under different user numbers.}
        \label{UE-UE}
    \end{center}
\end{figure}

\begin{figure}[!t]
    \begin{center}
        \includegraphics[width=0.9\columnwidth,height=2.5in]{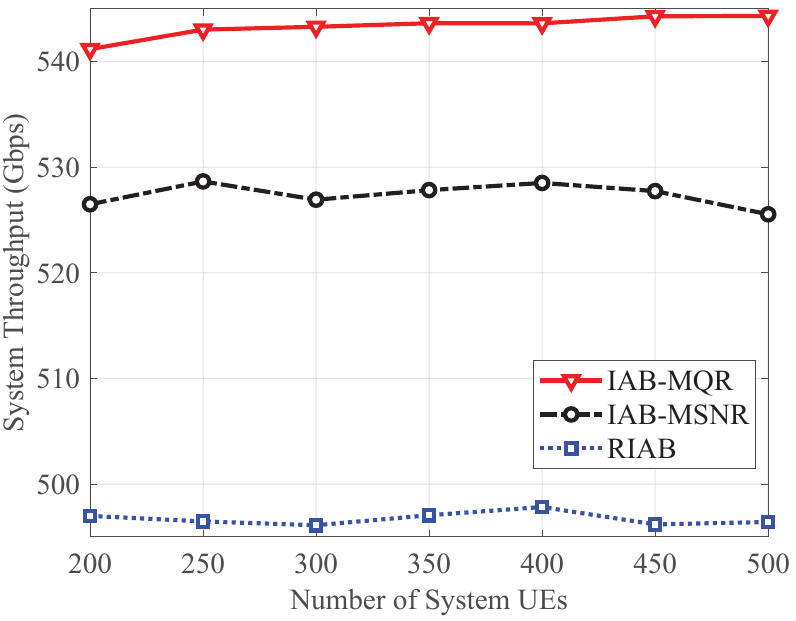}
        \caption{System throughput under different user numbers.}
        \label{UE-th}
    \end{center}
\end{figure}

\subsection{Comparison with Baseline Schemes}\label{S4-2}

In order to verify the effectiveness of the proposed algorithm in transmission scheduling, the algorithm is compared with the following baseline algorithms.

\begin{itemize}
\item {\em Integrated access and backhaul with maximum signal-to-noise (IAB-MSNR)}: The algorithm takes the SNR as the association criterion, that is, each UE is associated with the SBS providing the maximum SNR \cite{snr}. The backhaul scheduling is the same as the proposed algorithm, but the number of access transmission slots allocated to users is not updated.

\item {\em Random and integrated access and backhaul (RIAB)}: UEs will randomly connect to an SBS with this algorithm. In the backhaul scheduling phase, if the total number of slots required by the associated UEs is greater than the actual number of backhaul transmission slots $M$, the associted UEs are randomly eliminated.
\end{itemize}

The proposed algorithm is referred to as the \textit{IAB-MQR} algorithm in this algorithm.

The results of the number of served UEs and system throughput of the three algorithms under different user numbers are shown in Fig.~\ref{UE-UE} and Fig.~\ref{UE-th}, respectively. The transmit power of access links and backhaul links are both 1,000 mW, and the number of access transmission slots and the number of backhaul transmission slots are both 2,000. In Fig.~\ref{UE-UE}, as the number of UEs is increased, the number of served UEs of the proposed IAB-MQR algorithm is steadily rising, while the numbers of served users of the other two algorithms remain almost the same. Since the IAB-MQR algorithm takes into account the limited system resources, it preferentially selects and schedules the UEs with fewer required slots to serve more UEs, while the other two algorithms can only serve fewer UEs with limited resources. This also shows that the proposed algorithm can guarantee the service of more users and respond to a large number of bursty user requests under the condition of limited resources. When the number of system UEs is 500, the number of served UEs of the IAB-MQR algorithm is $35.5\%$ and $42.9\%$ higher than that of the IAB-MSNR algorithm and the RIAB algorithm, respectively.

In Fig.~\ref{UE-th}, as the number of system users grows up, the system throughput of the IAB-MQR algorithm fluctuates slightly around 543 Gbps, which shows the advantages of the proposed algorithm in improving network throughput. The IAB-MSNR algorithm selects UEs with the strongest SNR for association and transmission scheduling. UEs with high QoS requirements and low access rates occupy system resources for a long time, resulting in access failure of other UEs. Futhermore, the algorithm does not reallocate the slot resources of access links after backhaul scheduling, and the access transmission slots are not fully utilized. Similarly, the RIAB algorithm may have more UEs occupying system resources all the time than the IAB-MSNR algorithm, and the performance of this algorithm is the worst.

\begin{figure}[t!]
    \begin{center}
        \includegraphics[width=0.9\columnwidth,height=2.5in]{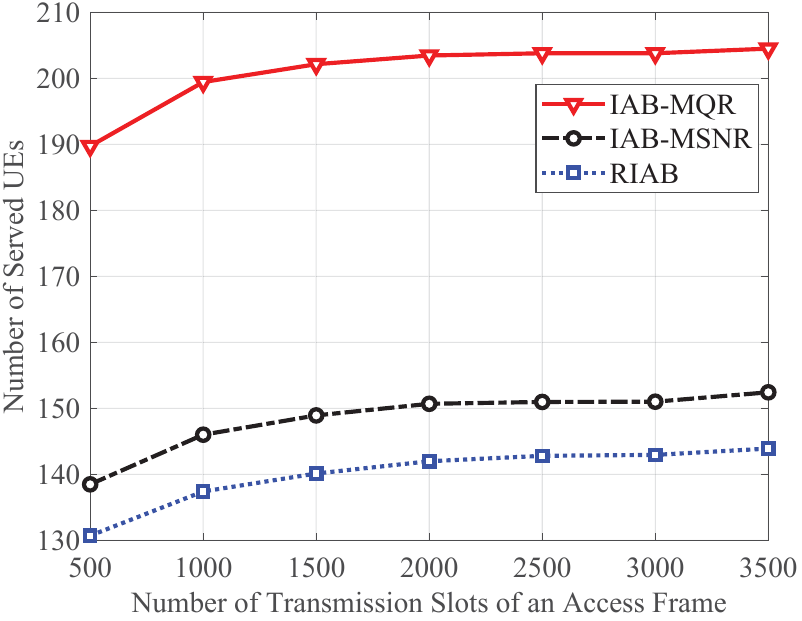}
        \caption{The number of served UEs under different transmission slots of an access frame.}
        \label{ac-ue}
    \end{center}
\end{figure}

\begin{figure}[!t]
    \begin{center}
        \includegraphics[width=0.9\columnwidth,height=2.5in]{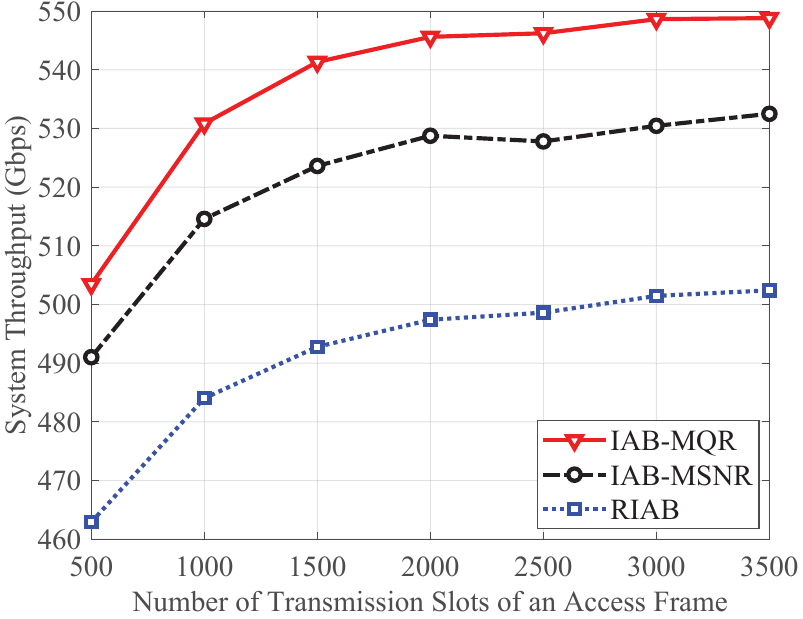}
        \caption{System throughput under different transmission slots of an access frame.}
        \label{ac-th}
    \end{center}
\end{figure}

\begin{figure}[!t]
    \begin{center}
        \includegraphics[width=0.9\columnwidth,height=2.5in]{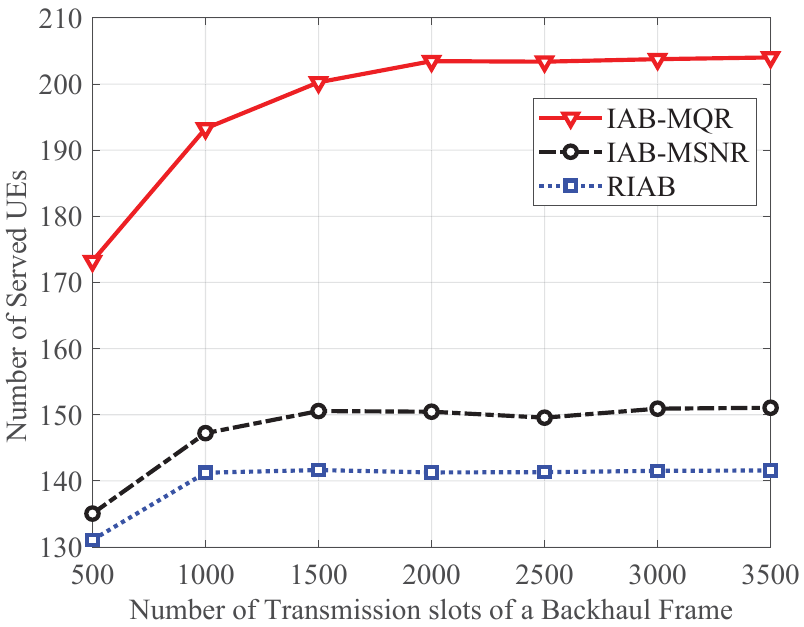}
        \caption{The number of served UEs under different transmission slots of a backhaul frame.}
        \label{ba-ue}
    \end{center}
\end{figure}

\begin{figure}[!t]
    \begin{center}
        \includegraphics[width=0.9\columnwidth,height=2.5in]{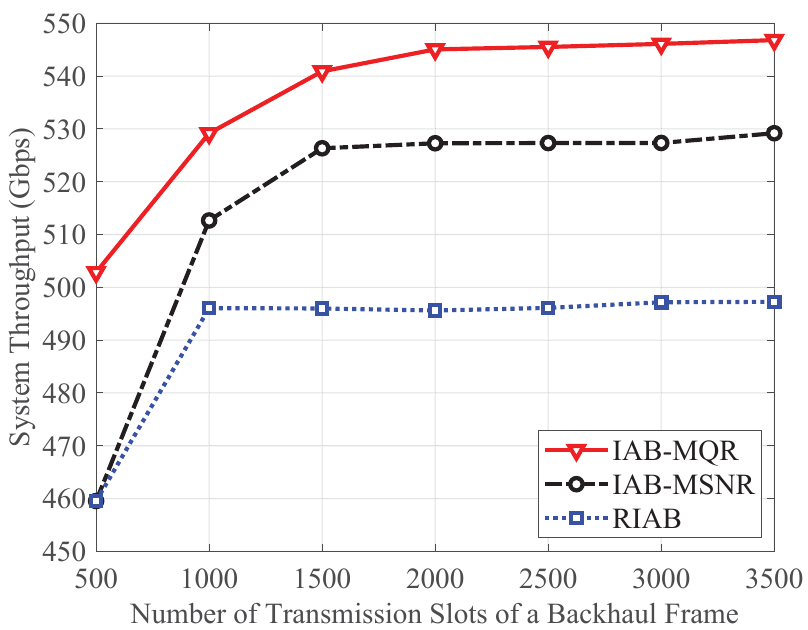}
        \caption{System throughput under different transmission slots of a backhaul frame.}
        \label{ba-th}
    \end{center}
\end{figure}

Fig.~\ref{ac-ue} and Fig.~\ref{ac-th} show the number of served UEs and system throughput with various access transmission slots, respectively. We set the transmit power of both access links and backhaul links to 1,000 mW, the number of backhaul transmission slots to 2,000, and the number of users to 500. The proposed algorithm is superior to the other two algorithms in terms of the number of served UEs and system throughput. With the increase of access transmission slots, the number of served UEs and system throughput of the three algorithms also increase, and the growth rate slows down. When there are fewer access transmission slots, there are fewer schedulable access links in the access phase, that is, fewer associated UEs, resulting in a portion of unused backhaul transmission slot resources and fewer schedulable backhaul links. As the number of access transmission slots goes up, so does the number of scheduled backhaul links and therefore the number of served UEs and system throughput. After the number of access transmission slots reaches 2,000, the number of served UEs of the proposed IAB-MQR algorithm remains stable. This is because under the constraints of backhaul slots, the scheduled backhaul links are also limited. It is worth noting that the system throughput of the proposed IAB-MQR algorithm still increases slightly after the number of access transmission slots reaches 2,000. The reason is that the algorithm updates the number of access slots allocated to UEs according to \eqref{updata} after the completion of backhaul scheduling, which makes full use of the access slot resources and further improves the actual throughput of access links. Because the system throughput satisfies $C = \sum\nolimits_\mathcal{U} {\left( {{x_{{b_l},{u_k}}} \cdot \min \left( {C_{{u_k}}^A,C_{{u_k}}^B} \right)} \right)}$, the system throughput also increases slightly.

\begin{figure}[t!]
    \begin{center}
        \includegraphics[width=0.9\columnwidth,height=2.5in]{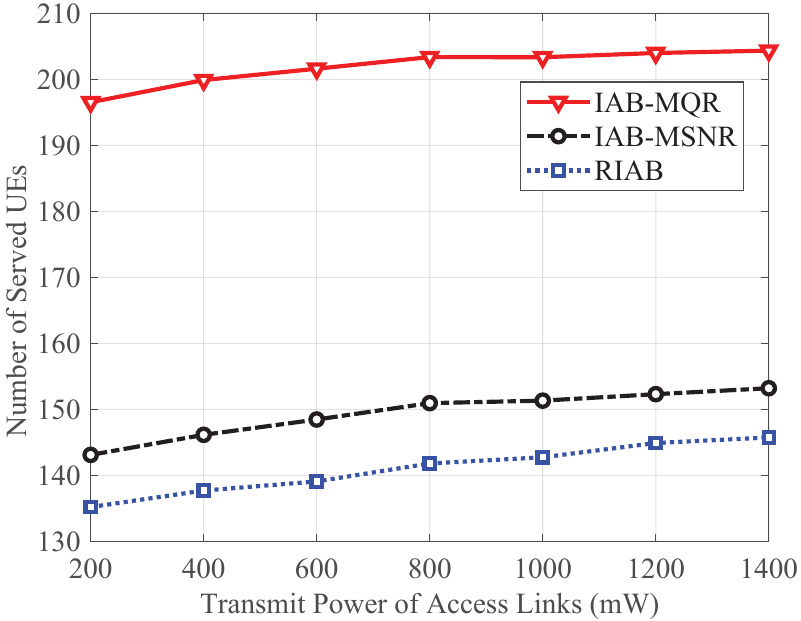}
        \caption{Number of served UEs under different transmit power of access links.}
        \label{pa-ue}
    \end{center}
\end{figure}

\begin{figure}[t!]
    \begin{center}
        \includegraphics[width=0.9\columnwidth,height=2.5in]{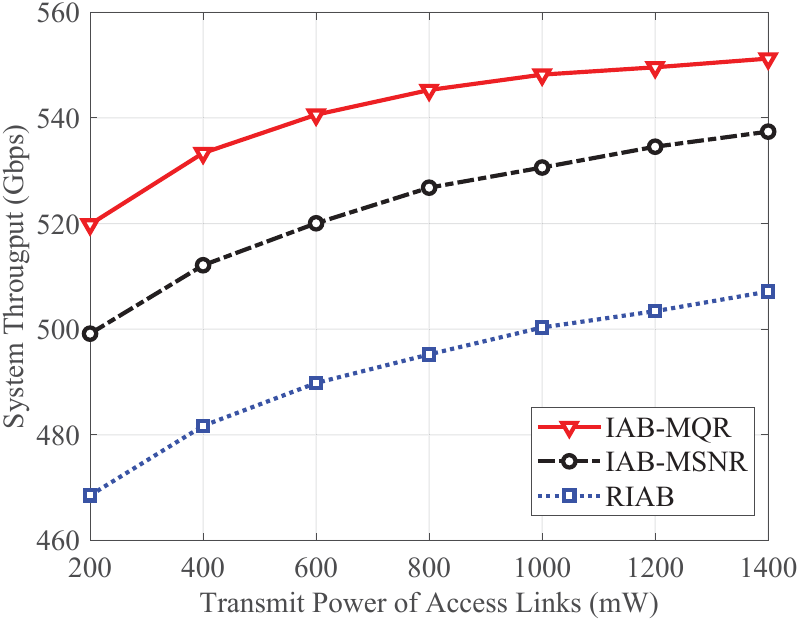}
        \caption{System throughput under different transmit power of access links.}
        \label{pa-th}
    \end{center}
\end{figure}

Fig.~\ref{ba-ue} and Fig.~\ref{ba-th} plot the performance comparison of the three schemes under different backhaul transmission slots, where the transmit power of access links and backhaul links is 1,000 mW, the number of access transmission slots is 2,000, and the number of users is 500. The proposed algorithm serves the most users and achieves the highest throughput. When the number of backhaul transmission slots is small, the QoS requirements of most users cannot be satisfied and thus there are fewer backhaul links that can be scheduled. In this case, the number of backhaul transmission slots is the main factor limiting the performance of the algorithm. Regarding system throughput, the IAB-MQR algorithm reallocates transmission slots of access links after backhaul scheduling, hence users achieves higher throughput on access links. However, the actual throughput depends on the minimum throughput achieved by the access link and the backhaul link, the throughput of the IAB-MQR algorithm is also limited by the finite backhaul transmission slots. In addition, when the number of backhaul transmission slots is 2,000, the performance of the proposed algorithm is not further improved, while the values of the other two algorithms are 1,500 and 1,000, respectively. This difference also shows the advantages of the proposed algorithm in resource allocation.

\begin{figure}[t!]
    \begin{center}
        \includegraphics[width=0.9\columnwidth,height=2.5in]{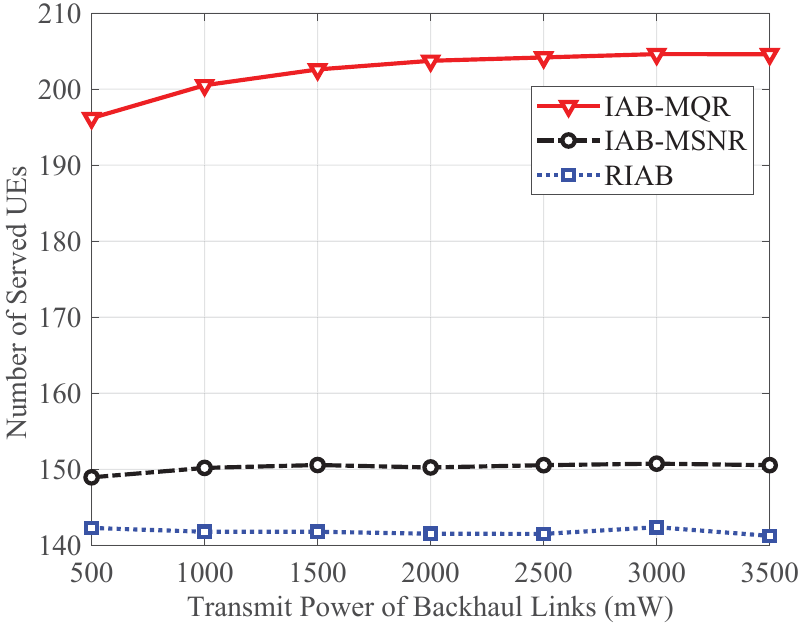}
        \caption{Number of served UEs under different transmit power of backhaul links.}
        \label{pb-ue}
    \end{center}
\end{figure}

\begin{figure}[t!]
    \begin{center}
        \includegraphics[width=0.9\columnwidth,height=2.5in]{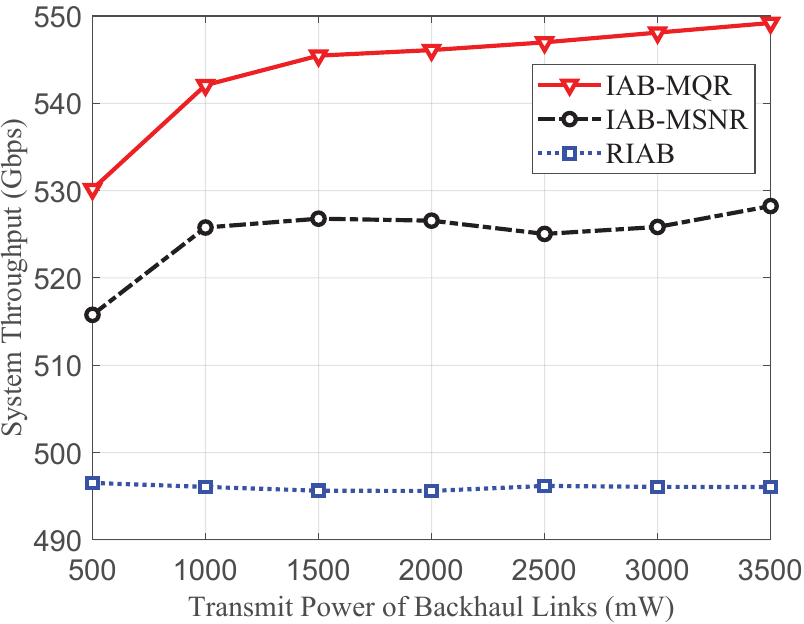}
        \caption{System throughput under different transmit power of backhaul links.}
        \label{pb-th}
    \end{center}
\end{figure}

From Fig.~\ref{ac-ue} to Fig.~\ref{ba-th}, it can be seen that setting the appropriate number of access transmission slots and backhaul transmission slots helps improve the overall performance of the network while making full use of system resources.

The number of served UEs and system throught with varying transmit power of access links are shown in Fig.~\ref{pa-ue} and Fig.~\ref{pa-th} with $P_T^B=1,000$ mW, $N=3,000$, $M=2,000$ and $K=500$. The proposed algorithm outperforms the other algorithms with respect to these two performance metrics. When the access transmission power is 1,400 mW, the number of served users of the IAB-MQR algorithm is still $33.4\%$ and $40.2\%$ higher than that of the IAB-MSNR algorithm and the RIAB algorithm, respectively. Particularly, when the access transmit power exceeds 800 mW, the number of served UEs of the proposed IAB-MQR algorithm remains nearly stable, and the system throughput still increases slightly. Higher access transmission power helps to improve the transmission rate of access links. Therefore, with the same number of access transmission slots, the access link can achieve a greater throughput at higher access transmission power. On the other hand, when the backhaul resources are consistent, the performance of the IAB-MQR algorithm is basically the same under the following two simulation settings: (i) the access transmit power is 800 mW, and the number of access transmission slots is 3,000, (ii) the access transmit power is 1,000 mW, and the number of access transmission slots is 2,000. This also shows that appropriately increasing the number of access transmission slots can reduce the requirements for access transmit power.

Fig.~\ref{pb-ue} and Fig.~\ref{pb-th} plot the results of the three algorithms under different transmit power of backhaul links with $P_T^A=800$ mW, $N=3,000$, $M=1,500$, and $K=500$. When the backhaul transmit power exceeds 2,000 mW, the number of served users of the IAB-MQR algorithm is almost constant, and its system throughput still goes up. Compared with Fig.~\ref{pa-ue} and Fig.~\ref{pa-th}, the curves of IAB-MSNR algorithm and RIAB algorithm are relatively flat in general. These two algorithms do not update the number of access slots of associated UEs after the backhaul scheduling. According to \eqref{C} and \eqref{t1}, the actual throughput of access links of the two algorithms is only slightly greater than the QoS requirements of users. Even if the transmit power of the backhaul link is increased, the actual throughput of the system cannot be increased.

\section{Conclusion}\label{S5}

Considering the high data rate requirements of users, this paper studied the transmission scheduling problem of integrated mmWave access and terahertz backhaul networks. In order to maximize the number of served users under the constraints of limited time resources and the half-duplex mode of BSs, we proposed a minimum rate ratio user association and IAB transmission scheduling algorithm. The BSs first allocate slots to the UE with the smallest rate ratio and make pre-decision of access scheduling. Then, in the backhaul scheduling phase, the algorithm preferentially schedules users with a small number of backhaul slots, and finally updates the access users according to the backhaul scheduling results. In particular, different QoS requirements of users were taken into account, which was more suitable for the actual network state. The proposed algorithm was evaluated under different parameters including the number of UEs, the number of access transmission slots, the number of backhaul transmission slots, and the transmit power of access links and backhaul links. The simulation results showed that the proposed algorithm outperforms other benchmark schemes in terms of the number of served users and system throughput. Moreover, setting the appropriate number of access and backhaul transmission slots helped to improve the network performance.

\bibliographystyle{IEEEtran}
\bibliography{mybib}

\begin{thebibliography}{10}
\providecommand{\url}[1]{#1}
\csname url@samestyle\endcsname
\providecommand{\newblock}{\relax}
\providecommand{\bibinfo}[2]{#2}
\providecommand{\BIBentrySTDinterwordspacing}{\spaceskip=0pt\relax}
\providecommand{\BIBentryALTinterwordstretchfactor}{4}
\providecommand{\BIBentryALTinterwordspacing}{\spaceskip=\fontdimen2\font plus
\BIBentryALTinterwordstretchfactor\fontdimen3\font minus
  \fontdimen4\font\relax}
\providecommand{\BIBforeignlanguage}[2]{{%
\expandafter\ifx\csname l@#1\endcsname\relax
\typeout{** WARNING: IEEEtran.bst: No hyphenation pattern has been}%
\typeout{** loaded for the language `#1'. Using the pattern for}%
\typeout{** the default language instead.}%
\else
\language=\csname l@#1\endcsname
\fi
#2}}
\providecommand{\BIBdecl}{\relax}
\BIBdecl

\bibitem{mmwave}
S.~He, Y.~Zhang, J.~Wang, J.~Zhang, J.~Ren, Y.~Zhang, W.~Zhuang, and X.~Shen,
  ``A survey of millimeter-wave communication: Physical-layer technology
  specifications and enabling transmission technologies,'' \emph{Proceedings of
  the IEEE}, vol. 109, no.~10, pp. 1666--1705, Oct. 2021.

\bibitem{udn}
M.~Filo, C.~H. Foh, S.~Vahid, and R.~Tafazolli, ``Performance analysis of
  ultra-dense networks with regularly deployed base stations,'' \emph{IEEE
  Transactions on Wireless Communications}, vol.~19, no.~5, pp. 3530--3545, May
  2020.

\bibitem{fiber}
X.~Ge, S.~Tu, G.~Mao, V.~K.~N. Lau, and L.~Pan, ``Cost efficiency optimization
  of {5G} wireless backhaul networks,'' \emph{IEEE Transactions on Mobile
  Computing}, vol.~18, no.~12, pp. 2796--2810, Dec. 2019.

\bibitem{iab}
M.~Gupta, A.~Rao, E.~Visotsky, A.~Ghosh, and J.~G. Andrews, ``Learning link
  schedules in self-backhauled millimeter wave cellular networks,'' \emph{IEEE
  Transactions on Wireless Communications}, vol.~19, no.~12, pp. 8024--8038,
  Dec. 2020.

\bibitem{6g}
M.~Z. Chowdhury, M.~Shahjalal, S.~Ahmed, and Y.~M. Jang, ``{6G} wireless
  communication systems: Applications, requirements, technologies, challenges,
  and research directions,'' \emph{IEEE Open Journal of the Communications
  Society}, vol.~1, pp. 957--975, 2020.

\bibitem{intro1}
S.~Ghafoor, N.~Boujnah, M.~H. Rehmani, and A.~Davy, ``{MAC} protocols for
  terahertz communication: A comprehensive survey,'' \emph{IEEE Communications
  Surveys Tutorials}, vol.~22, no.~4, pp. 2236--2282, Fourthquarter 2020.

\bibitem{intro2}
V.~Petrov, J.~Kokkoniemi, D.~Moltchanov, J.~Lehtomaki, Y.~Koucheryavy, and
  M.~Juntti, ``Last meter indoor terahertz wireless access: Performance
  insights and implementation roadmap,'' \emph{IEEE Communications Magazine},
  vol.~56, no.~6, pp. 158--165, June 2018.

\bibitem{intro3}
J.~Wang, C.-X. Wang, J.~Huang, H.~Wang, and X.~Gao, ``A general {3D}
  space-time-frequency non-stationary {THz} channel model for {6G}
  ultra-massive {MIMO} wireless communication systems,'' \emph{IEEE Journal on
  Selected Areas in Communications}, vol.~39, no.~6, pp. 1576--1589, June 2021.

\bibitem{sara}
A.~S. Cacciapuoti, K.~Sankhe, M.~Caleffi, and K.~R. Chowdhury, ``Beyond {5G}:
  {THz}-based medium access protocol for mobile heterogeneous networks,''
  \emph{IEEE Communications Magazine}, vol.~56, no.~6, pp. 110--115, June 2018.

\bibitem{ua1}
Y.~Sun, G.~Feng, S.~Qin, and S.~Sun, ``Cell association with user behavior
  awareness in heterogeneous cellular networks,'' \emph{IEEE Transactions on
  Vehicular Technology}, vol.~67, no.~5, pp. 4589--4601, May 2018.

\bibitem{ua3}
R.~Liu, Q.~Chen, G.~Yu, and G.~Y. Li, ``Joint user association and resource
  allocation for multi-band millimeter-wave heterogeneous networks,''
  \emph{IEEE Transactions on Communications}, vol.~67, no.~12, pp. 8502--8516,
  Dec. 2019.

\bibitem{ua4}
X.~Ge, X.~Li, H.~Jin, J.~Cheng, and V.~C.~M. Leung, ``Joint user association
  and user scheduling for load balancing in heterogeneous networks,''
  \emph{IEEE Transactions on Wireless Communications}, vol.~17, no.~5, pp.
  3211--3225, May 2018.

\bibitem{ua2}
N.~Zhao, Y.-C. Liang, D.~Niyato, Y.~Pei, M.~Wu, and Y.~Jiang, ``Deep
  reinforcement learning for user association and resource allocation in
  heterogeneous cellular networks,'' \emph{IEEE Transactions on Wireless
  Communications}, vol.~18, no.~11, pp. 5141--5152, Nov. 2019.

\bibitem{thzb1}
C.-X. Wang, J.~Wang, S.~Hu, Z.~H. Jiang, J.~Tao, and F.~Yan, ``Key technologies
  in {6G} terahertz wireless communication systems: A survey,'' \emph{IEEE
  Vehicular Technology Magazine}, vol.~16, no.~4, pp. 27--37, Dec. 2021.

\bibitem{thzloss}
A.~Shafie, N.~Yang, S.~Durrani, X.~Zhou, C.~Han, and M.~Juntti, ``Coverage
  analysis for {3D} terahertz communication systems,'' \emph{IEEE Journal on
  Selected Areas in Communications}, vol.~39, no.~6, pp. 1817--1832, June 2021.

\bibitem{thzb2}
H.~Yuan, N.~Yang, X.~Ding, C.~Han, K.~Yang, and J.~An, ``Cluster-based
  multi-carrier hybrid beamforming for massive device terahertz
  communications,'' \emph{IEEE Transactions on Communications}, vol.~70, no.~5,
  pp. 3407--3420, May 2022.

\bibitem{thz3}
P.~Bhardwaj and S.~M. Zafaruddin, ``Performance of dual-hop relaying for
  {THz-RF} wireless link,'' in \emph{Proc. 2021 IEEE 93rd Vehicular Technology
  Conference}, Helsinki, Finland, Apr. 2021, pp. 1--5.

\bibitem{thz1}
M.~Yu, A.~Tang, X.~Wang, and C.~Han, ``Joint scheduling and power allocation
  for {6G} terahertz mesh networks,'' in \emph{Proc. 2020 International
  Conference on Computing, Networking and Communications}, Big Island, HI, USA,
  Feb. 2020, pp. 631--635.

\bibitem{ra2}
C.~Saha and H.~S. Dhillon, ``Millimeter wave integrated access and backhaul in
  {5G}: Performance analysis and design insights,'' \emph{IEEE Journal on
  Selected Areas in Communications}, vol.~37, no.~12, pp. 2669--2684, Dec.
  2019.

\bibitem{ra3}
W.~Lei, Y.~Ye, and M.~Xiao, ``Deep reinforcement learning-based spectrum
  allocation in integrated access and backhaul networks,'' \emph{IEEE
  Transactions on Cognitive Communications and Networking}, vol.~6, no.~3, pp.
  970--979, Sept. 2020.

\bibitem{ra4}
A.~S. Tan and E.~D. Biyar, ``{QoS}-aware autonomous {IAB}-node activation and
  access control,'' in \emph{Proc. 2021 IEEE International Conference on
  Communications Workshops}, Montreal, QC, Canada, July 2021, pp. 1--6.

\bibitem{ra5}
M.~Pagin, T.~Zugno, M.~Polese, and M.~Zorzi, ``Resource management for {5G}
  {NR} integrated access and backhaul: A semi-centralized approach,''
  \emph{IEEE Transactions on Wireless Communications}, vol.~21, no.~2, pp.
  753--767, Feb. 2022.

\bibitem{thzband2}
C.~Han, A.~O. Bicen, and I.~F. Akyildiz, ``Multi-wideband waveform design for
  distance-adaptive wireless communications in the terahertz band,'' \emph{IEEE
  Transactions on Signal Processing}, vol.~64, no.~4, pp. 910--922, Feb. 2016.

\bibitem{mmband}
A.~N. Uwaechia and N.~M. Mahyuddin, ``A comprehensive survey on millimeter wave
  communications for fifth-generation wireless networks: Feasibility and
  challenges,'' \emph{IEEE Access}, vol.~8, pp. 62\,367--62\,414, Mar. 2020.

\bibitem{thzband}
K.~M.~S. Huq, S.~A. Busari, J.~Rodriguez, V.~Frascolla, W.~Bazzi, and D.~C.
  Sicker, ``Terahertz-enabled wireless system for beyond-{5G} ultra-fast
  networks: A brief survey,'' \emph{IEEE Network}, vol.~33, no.~4, pp. 89--95,
  July 2019.

\bibitem{gbps}
Z.~Zhang, Y.~Xiao, Z.~Ma, M.~Xiao, Z.~Ding, X.~Lei, G.~K. Karagiannidis, and
  P.~Fan, ``{6G} wireless networks: Vision, requirements, architecture, and key
  technologies,'' \emph{IEEE Vehicular Technology Magazine}, vol.~14, no.~3,
  pp. 28--41, Sept. 2019.

\bibitem{antenna}
Q.~Chen, X.~Peng, J.~Yang, and F.~Chin, ``Spatial reuse strategy in {mmWave}
  {WPANs} with directional antennas,'' in \emph{Proc. 2012 IEEE Global
  Communications Conference}, Anaheim, CA, Dec. 2012, pp. 5392--5397.

\bibitem{antenna2}
H.~Sawada, A.~Kanno, N.~Yamamoto, K.~Fujii, A.~Kasamatsu, K.~Ishizu, F.~Kojima,
  H.~Ogawa, and I.~Hosako, ``High gain antenna characteristics for 300 {GHz}
  band fixed wireless communication systems,'' in \emph{Proc. 2017 Progress in
  Electromagnetics Research Symposium - Fall}, Singapore, Nov. 2017, pp.
  1409--1412.

\bibitem{model}
J.~M. Jornet and I.~F. Akyildiz, ``Channel modeling and capacity analysis for
  electromagnetic wireless nanonetworks in the terahertz band,'' \emph{IEEE
  Transactions on Wireless Communications}, vol.~10, no.~10, pp. 3211--3221,
  Oct. 2011.

\bibitem{rothman2009hitran}
L.~S. Rothman, I.~E. Gordon, A.~Barbe, D.~C. Benner, P.~F. Bernath, M.~Birk,
  V.~Boudon, L.~R. Brown, A.~Campargue, J.-P. Champion \emph{et~al.}, ``The
  {HITRAN} 2008 molecular spectroscopic database,'' \emph{Journal of
  Quantitative Spectroscopy and Radiative Transfer}, vol. 110, no. 9-10, pp.
  533--572, June 2009.

\bibitem{snr}
J.~G. Andrews, ``Seven ways that {HetNets} are a cellular paradigm shift,''
  \emph{IEEE Communications Magazine}, vol.~51, no.~3, pp. 136--144, Mar. 2013.

\end{thebibliography}

\end{document}